\documentclass[aps,prd,twocolumn,floatfix,altaffilletter,superscriptaddress,preprintnumbers,
tightenlines,showpacs,showkeys,nofootinbib,notoccite]{revtex4-1}
\usepackage[utf8]{inputenc}
\usepackage[colorlinks=true,citecolor=blue,linkcolor=blue]{hyperref}
\usepackage[normalem]{ulem}
\usepackage{amsmath,amssymb, mathrsfs,esvect}
\usepackage{epsfig}
\usepackage{graphicx}               % Standard graphics package
\usepackage{url}
\usepackage{color}
\usepackage{slashed}
\usepackage{multirow}
\usepackage{placeins}
\usepackage[dvipsnames]{xcolor}
\usepackage{epstopdf}
\usepackage{siunitx}
\usepackage{soul}
\usepackage{tikz}
\usepackage[capitalise, english]{cleveref}
\usepackage{siunitx}
\usepackage{xspace}
\usepackage{booktabs}
\usepackage[capitalise]{cleveref}
\usetikzlibrary{trees}
\usetikzlibrary{decorations.pathmorphing}
\usetikzlibrary{decorations.markings}

\newcommand\myshade{80}
\colorlet{mylinkcolor}{ForestGreen}
\colorlet{mycitecolor}{Red}
\colorlet{myurlcolor}{violet}

\hypersetup{
  pdfauthor={U.B,, N.R., A.R.},
  pdftitle={Happy Dvali},
  pdfsubject={Happy Dvali}
	linkcolor  = mylinkcolor!\myshade!black,
	citecolor  = mycitecolor!\myshade!black,
	urlcolor   = myurlcolor!\myshade!black,
	colorlinks = true
}

\definecolor{jblue}{RGB}{20,50,100}
\definecolor{npurple}{RGB} {153, 51, 204}
\definecolor{wred}{RGB}{217,0,56}
\definecolor{white}{RGB}{255,255,255}
\definecolor{forestgreen}{HTML}{228B22}

\allowdisplaybreaks
\usepackage{tikz,xcolor,hyperref}

\definecolor{lime}{HTML}{A6CE39}
\DeclareRobustCommand{\orcidicon}{\hspace{-1mm}
	\begin{tikzpicture}
		\draw[lime, fill=lime] (0,0) 
		circle [radius=0.16] 
		node[white] {{\fontfamily{qag}\selectfont \tiny \,ID}};
		\draw[white, fill=white] (-0.0525,0.095) 
		circle [radius=0.007];
	\end{tikzpicture}
	\hspace{-3mm}
}

\foreach \x in {A, ..., Z}{\expandafter\xdef\csname orcid\x\endcsname{\noexpand\href{https://orcid.org/\csname orcidauthor\x\endcsname}
		{\noexpand\orcidicon}}
}

\definecolor{aquamarine}{rgb}{0.2,0.7,0.6}
\definecolor{cerulean}{RGB}{0,166,214} 
\definecolor{subtlered}{rgb}{0.8,0.3,0.3}

\begin{document}
%%%%%%%%%%%%%%%%%

\title{Beyond Hawking evaporation of black holes \\ formed by dark matter in compact stars}
	\author{Ujjwal Basumatary\orcidA}
	\email{ujjwalb@iisc.ac.in}
	\affiliation{Centre for High Energy Physics, Indian Institute of Science, C. V. Raman Avenue, Bengaluru 560012, India}
	\author{Nirmal Raj\orcidB}
	\email{nraj@iisc.ac.in}
		\affiliation{Centre for High Energy Physics, Indian Institute of Science, C. V. Raman Avenue, Bengaluru 560012, India}
	\author{Anupam Ray\orcidD}
	\email{anupam.ray@berkeley.edu}
 \affiliation{Department of Physics, University of California Berkeley, Berkeley, California 94720, USA}
\date{\today}
	
%%%%%%%%%%%%%%%%%%%%%%%%%%%%%%%%%%%%%%%%%%%%%%%%%%%%%%%%%%%%%%%%%%%
\begin{abstract}
The memory burden effect is an explicit resolution to the information paradox by which an evaporating black hole acquires quantum hair, which then suppresses its rate of mass loss with respect to the semi-classical Hawking rate.
We show that this has significant implications for particle dark matter that captures in neutron stars and forms black holes that go on to consume the host star. 
In particular, we show that constraints on the nucleon scattering cross section and mass of spin-0 and spin-1/2 dark matter would be extended by several orders of magnitude.
\end{abstract}
\maketitle

\preprint{N3AS-24-036}

%%%
\section{Introduction}
\label{sec:intro}
%%%%
%%%
\begin{figure*}[t]
    \centering
    \includegraphics[width=0.45\textwidth]{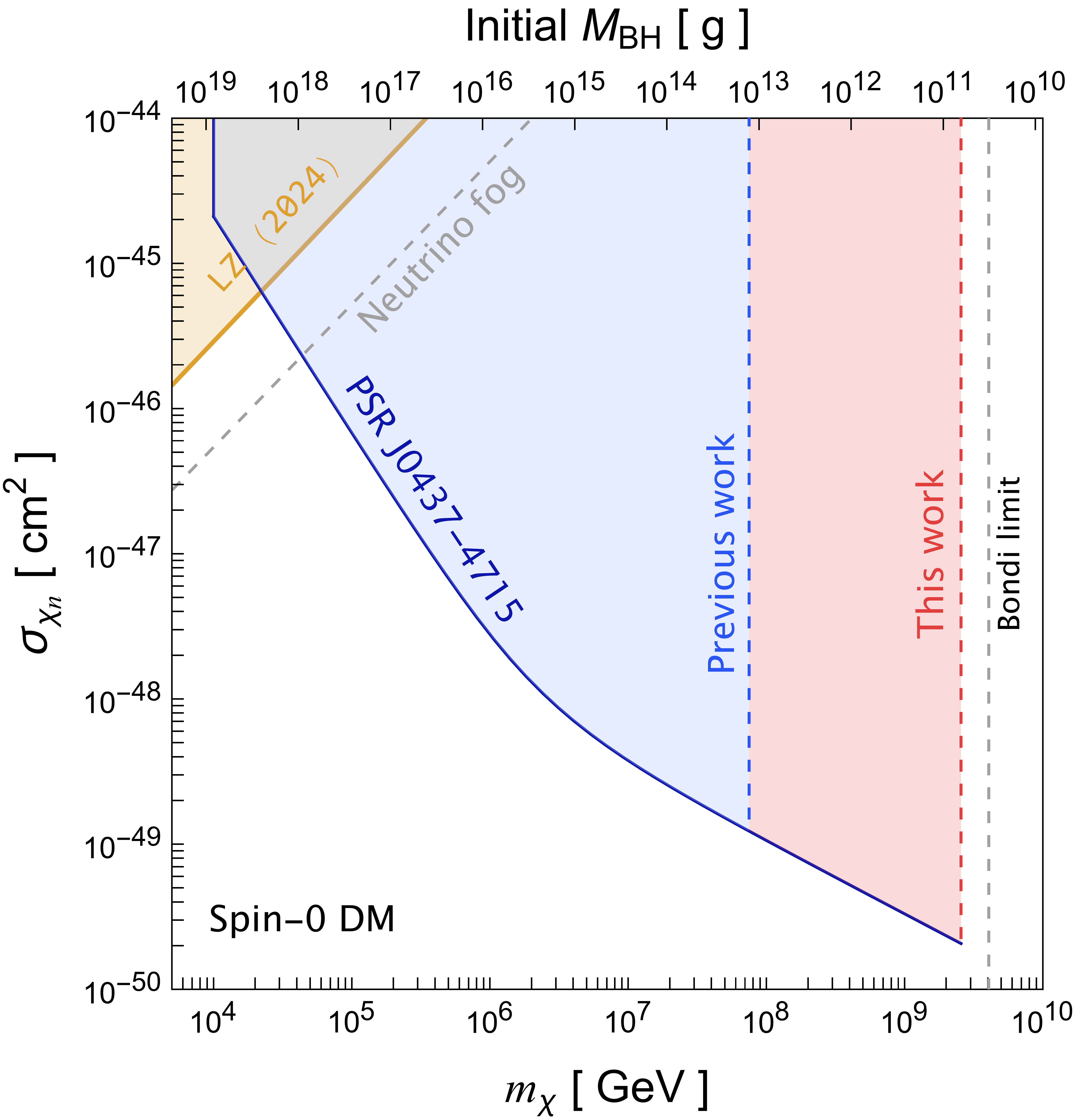}
    \includegraphics[width=0.45\textwidth]{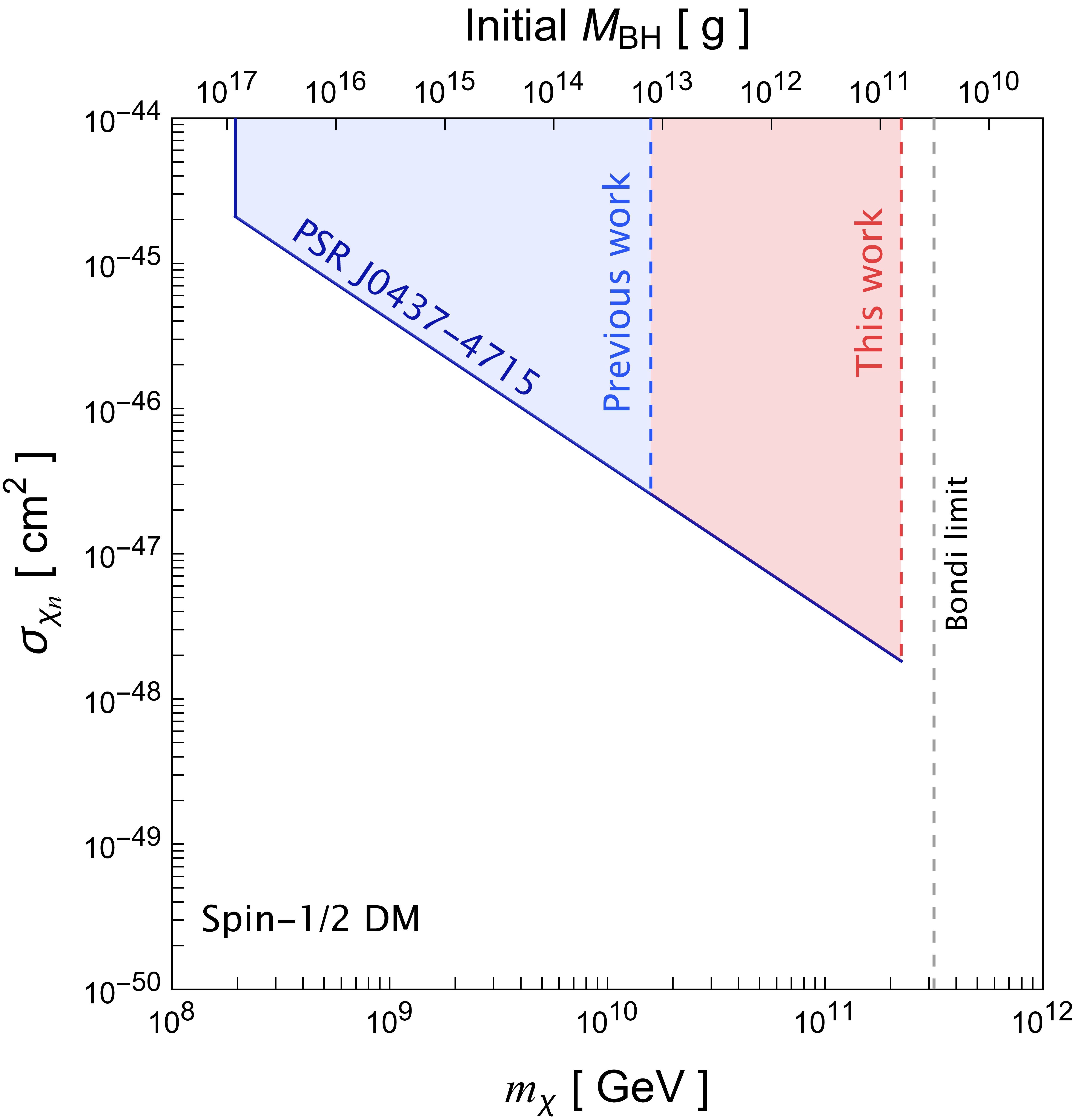}
    \caption{Limits on the dark matter-neutron scattering cross section as a function of dark matter mass from the survival of the neutron star PSR~J0437$-$4715; the top x-axis shows the initial mass of dark matter-seeded black holes corresponding to the bottom x-axis, as obtained from Eqs.~\eqref{eq:Mselfgrav} and \eqref{eq:MChandra}. 
    The red shaded region shows our extension of previous limits, namely, the blue shaded region, when suppressed evaporation of these black holes via the memory burden effect is taken into account.
    The gray vertical line shows the dark matter mass above which the neutron star will always survive due to the black hole Bondi-accreting stellar material on a timescale smaller than the age of the star.
    Also shown for comparison are recent constraints from a direct search at LZ~\cite{LZCollaboration:2024lux} and the neutrino fog~~\cite{OHare:2021utq}, obtained by linearly extrapolating exclusion cross sections at high dark matter masses. We have assumed here that half the mass of the black hole is lost before the memory burden phase, and that the exponent $k=1$ in Eq.~\eqref{eq:MevapdotMB}, but altering these choices have little effect on our limits.}
    \label{fig:limits}
\end{figure*}
%%%
How do black holes evaporate?
Resolutions to the black hole information paradox generically suggest that the semi-classical treatment of Hawking radiation breaks down beyond some point as a black hole sheds entropy~\cite{Dvali:2015aja}.
In this {\em Letter} we show that this observation has wide-reaching implications for the discovery of {\em particle} dark matter.
Halo dark matter interacting with nucleons may capture in stellar bodies, thermalize with their dense interior, and sink and collect in a region compact enough to form a small central black hole~\cite{Goldman:1989nd}.
If this black hole evaporates at a rate slower than that of accretion of the surrounding material, it would transmute the host star into a stellar mass black hole;
thus observations of the existence of celestial bodies constrain the nature of dark matter interactions.
Previous work had always assumed Hawking evaporation rates to derive these limits.
We will relax this assumption and consider models of slower evaporation, and to illustrate our point focus on the densest stellar objects, neutron stars.
The resulting new limits are summarized in Fig.~\ref{fig:limits}.

The suppression of a black hole's evaporation at the end stages of its decay via Hawking radiation is predicted by several effects, such as when black holes are magnetically charged~\cite{magBH:Stojkovic:2004hz,magBH:Maldacena:2020skw,magBH:Estes:2022buj,magBH:banerjee2024}, attain extremality near the Planck mass~\cite{extremBH:Chen:2002tu,extremBH:Alexeyev:2002tg,extremBH:Chen:2004ft,extremBH:Nozari:2005ah,extremBH:Lehmann:2019zgt,extremBH:Profumo:2024fxq}, or quasi-extremality~\cite{quasiextremBH:deFreitasPacheco:2023hpb}, or form massive remnants~\cite{Giddings:1992hh}. 
Recently, Refs.~\cite{Dvali:2018xpy,Dvali:2020wft,Dvali:2024hsb} proposed a mechanism by which the breakdown of the Hawking treatment occurs at much earlier times, at the latest when the black hole has shed half its mass, i.e. around the Page time.
The basic principle is that the very large entropy of a black hole corresponds to a very high capacity for storage of information, which is possible through gapless degrees of freedom that can be arranged at a low energy cost. 
The initial stages of a black hole's evaporation could then be modeled semi-classically as in the Hawking picture, however, the information remaining within the black hole produces a backreaction on the black hole itself.
The backreaction acts to increase the gaps in the energy levels, and quantum effects can no longer be neglected in modeling the radiation, which is now suppressed.
Therefore the evaporation is no longer self-similar until the black hole shrinks to the extremal scale -- now it {\em does} have memory.
This ``memory burden effect" is generic to any decaying system with high information storage capacity, e.g. saturons~\cite{Dvali:2024hsb}.
It provides an explicit mechanism for release of information from a black hole after the Page time, thereby resolving the information paradox, and gives it a quantum hair that is not constrained by classical no-hair theorems. 
The phenomenological consequences of slow evaporation of black holes after half-decay have recently been used to derive novel limits on a number of scenarios, including primordial black holes as dark matter~\cite{Alexandre:2024nuo,thoss2024breakdownhawkingevaporationopens,Balaji:2024hpu,Haque:2024eyh,Dvali:2024hsb,Barman:2024iht,Bhaumik:2024qzd,Barman:2024ufm,Kohri:2024qpd,Zantedeschi:2024ram,Chianese:2024rsn,Barker:2024mpz,Borah:2024bcr,Loc:2024qbz}. 
In this work, we use this framework for the \textit{first} time to constrain scenarios involving particle dark matter.

Neutron stars, by virtue of their high densities, steep gravitational potentials and other unique properties, serve as extremely sensitive laboratories of dark matter~\cite{bramante2023dark}.
Broadly, they may aid in the discovery of particle dark matter either by signatures of heating~~\cite{Kouvaris:2007ay,Bertone:2007ae,Baryakhtar:2017dbj,Raj:2017wrv,Chen:2018ohx,Bell:2018pkk,Acevedo:2019agu,Garani:2019fpa,Hamaguchi:2019oev,Bell:2019pyc,Joglekar:2019vzy,Joglekar:2020liw,Bell:2020jou,Bell:2020lmm,Dasgupta:2020dik,Coffey:2022eav,NSheat:DarkBary:McKeen:2020oyr, NSheat:Mirror:McKeen:2021jbh,Bramante:2021dyx,Fujiwara:2022uiq,Chatterjee:2022dhp,Gresham:2022biw,Raj:2023azx,Alvarez:2023fjj,Raj:2024kjq,Bell:2024qmj,Ema:2024wqr,Baryakhtar:2022hbu,Carney:2022gse}, or the formation of black holes in their interior that may destroy them~\cite{Goldman:1989nd,Gould:1989gw,Bertone:2007ae,deLavallaz:2010wp,McDermott:2011jp,Kouvaris:2010jy,Kouvaris:2011fi,Bell:2013xk,Guver:2012ba,Bramante:2013hn,Bramante:2013nma,Kouvaris:2013kra,Bramante:2014zca,Garani:2018kkd,Kouvaris:2018wnh,Dasgupta:2020dik,Lin:2020zmm,Dasgupta:2020mqg,Goldman:1989nd,Gould:1989gw,Bertone:2007ae,deLavallaz:2010wp,McDermott:2011jp,Kouvaris:2010jy,Kouvaris:2011fi,Bell:2013xk,Guver:2012ba,Bramante:2013hn,Bramante:2013nma,Kouvaris:2013kra,Bramante:2014zca,Fuller:2014rza,Bramante:2017ulk,Garani:2018kkd,Kouvaris:2018wnh,Dasgupta:2020dik,Lin:2020zmm,Dasgupta:2020mqg,Takhistov:2020vxs,Garani:2021gvc,Garani:2022quc,Steigerwald:2022pjo,Bhattacharya:2023stq,Liang:2023nvo,Bramante:2024idl,Baryakhtar:2022hbu}.
The latter effect can occur in wider ranges of model parameters if the evaporation of the black hole is slowed down. In this work, we will take the memory burden effect as a concrete mechanism of suppressed, beyond-Hawking evaporation, and derive revised limits on dark matter interactions. The following sections estimate this effect in detail.

%%%%%%%%%%%%%%%%%%%%%%%%%%%%%%%%%%%%%%%%%%%
\section{Dark matter-seeded black holes in neutron stars}
%%%%%%%%%%%%%%%%%%%%%%%%%%%%%%%%%%%%%%%%%%%

In this section we briefly review the various stages involved in forming a black hole inside neutron stars via dark matter capture, and detail how the usual treatment of Hawking evaporation is modified by the memory burden effect. 
For more details on the black hole formation process see Ref.~\cite{bramante2023dark}.
As a benchmark for illustration and setting limits, we take the neutron star PSR~J0437$-$4715, with
mass 1.44\,$M_{\odot}$, 
radius 10.6 km, 
core density $1.4 \times 10^{15}$ g/cm$^3$, 
core temperature $2.1 \times 10^6$~K, 
and age 6.7~Gyr~\cite{Manchester:2004bp,McDermott:2011jp,Garani:2018kkd}.
%%%%%%
\subsection{Capture, thermalization, collapse, black hole formation}
%%%%%%

We assume that dark matter of mass $m_\chi$ has negligible or no self-annihilations in order to retain its captured population, that it carries a spin of either 0 or 1/2, and that its interaction with nucleons is velocity-independent. 
Then the rate of capture in a neutron star in the optically thin regime for dark matter-nucleon scattering cross section $\sigma_{\chi n}$  is~\cite{McDermott:2011jp,Bhattacharya:2023stq}
\begin{eqnarray}
 C_\chi &=& 10^{20} \, \text{s}^{-1} \left( \frac{\rho_\chi}{0.3 \, \text{GeV}/\text{cm}^{3}} \right) \left( \frac{10^5 \, \text{GeV}}{m_\chi} \right)  \\ \nonumber &\times& \left( \frac{\sigma_{\chi n}}{10^{-45} \, \text{cm}^2} \right) \left( \frac{220 \, \text{km}/ \text{s}}{\bar{v}_{\text{gal}}} \right) \left( 1 - \frac{1 - e^{-A^2}}{A^2} \right)  \,,
\\ 
\nonumber A^2 &=& \frac{6 m_\chi m_n v_{\text{esc}}^2}{\bar{v}_{\text{gal}}^2 (m_\chi - m_n)^2}~, 
\end{eqnarray}
where $\rho_{\chi}$ and $\bar{v}_{\rm gal}$ are normalized to the ambient density and average Maxwell-Boltzmann speed of dark matter particles, and the $A^2$ factor accounts for inefficient momentum transfers at large $m_\chi$~\cite{Bramante:2017xlb, Dasgupta:2019juq}.
 
Upon capture, the dark matter loses energy via repeated scattering and eventually reaches thermal equilibrium with the neutron star core of temperature $T_{\rm NS}$ over a timescale~\cite{Bertoni:2013bsa,Garani:2020wge}
\begin{equation}
    t_{\rm th} = 17~{\rm yr} \, \frac{m_\chi/m_n}{(1+m_\chi/m_n)^2} \left(\frac{10^{-45}~\rm cm^2}{\sigma_{\chi n}}\right)\left(\frac{2 \times 10^6 \, \rm K}{T_{\rm NS}}\right)^2\,.
\end{equation}
The captured and thermalized dark matter sinks downward and collects at the center of the star predominantly in a thermal sphere of radius~\cite{McDermott:2011jp}
\begin{multline}
    r_{\rm th} \simeq 0.03\,{\rm cm} \, \left(\frac{10^9~\rm GeV}{m_{\chi}}\right)^{1/2} \left(\frac{T_{\rm NS}}{2.1 \times 10^6 \,\rm K}\right)^{1/2}\\\times\left(\frac{1.4 \times 10^{15} \,\rm g/cm^3}{\rho_{\rm NSc}}\right)^{1/2}\,,
\end{multline}
where $\rho_{\rm NSc}$ is the neutron star central density. 
If the dark matter density within the thermal radius = $3 m_\chi N_\chi/(4\pi r^3_{\rm th})$ exceeds $\rho_{\rm NSc}$, the dark matter self-gravitates.
Consequently, a Jeans instability is initiated, and the dark matter sphere collapses.
The minimum accumulated mass of dark matter required for this to occur is~\cite{McDermott:2011jp, Kouvaris:2010jy} 
\begin{multline}
    M_\chi^{\rm self} = 1.5 \times 10^{35}~{\rm GeV} \left(\frac{10^9 \, \rm GeV}{m_\chi}\right)^{3/2}\left(\frac{T_{\rm NSc}}{2.1 \times 10^6\, {\rm K}}\right)^{3/2} \\ \times \left(\frac{1.4 \times 10^{15} \,\rm g/cm^3}{\rho_{\rm NSc}}\right)^{1/2}\,.
    \label{eq:Mselfgrav}
\end{multline}
To form a black hole we also require that the dark matter mass accumulated exceed its Chandrasekhar limit, given by~\cite{McDermott:2011jp,Kouvaris:2010jy}
%%%%%%
\begin{equation}
    M_{\chi}^{\rm Ch} = 
    \begin{cases}
     9.5 \times 10^{28}~{\rm GeV} \left(\frac{10^9 \, \rm GeV}{m_{\chi}}\right)\,~,~~{\rm spin}~0\,,\\
    1.8 \times 10^{39}~{\rm GeV} \left(\frac{10^9 \, \rm GeV}{m_{\chi}}\right)^2\,,~{\rm spin}~1/2~,
   \label{eq:MChandra}
  \end{cases}
\end{equation}
%%%%%
where for the spin-0 dark matter we have assumed no self-interactions.
A black hole is only formed after the mass captured over a time $t$, which is $m_\chi C_\chi t$, reaches max[$M_\chi^{\rm self}$, $M_\chi^{\rm Ch}$]. For our parametric range of interest, black hole formation for spin-0 dark matter is set by the self-gravitation criterion, and for spin-1/2 dark matter by the Chandrasekhar criterion.
In principle a small fraction of the nascent mass of the black hole may be emitted as gravitational radiation at the time of black hole formation~\cite{East:2019dxt}; we assume for simplicity that this is avoided by spherically symmetric collapse of dark matter.

The mass of the nascent black hole now evolves according to 
\begin{equation}
    \dot{M}_{\rm BH} = \frac{\pi \rho_{\rm NSc} G^2 M_{\rm BH}^2}{c_s^3} - \dot{M}_{\rm evap} \,,
    \label{eq:BHevolution}
\end{equation}
where the first term on the right hand side describes speherical Bondi-Hoyle accretion of the neutron star material with the sound speed $c_s = 0.17c$, and the second term is the evaporation rate.
The usual Hawking rate is
%%%%
\begin{equation}
\dot{M}^{\rm Hawk}_{\rm evap} = \frac{P(M_{\text{BH}})}{G^2 M_{\rm BH}^2}~,
\label{eq:Hawkrate}
\end{equation}
%%%%
where $P(M_{\text{BH}})$ is the Page factor accounting for gray-body corrections of the Hawking evaporation spectrum, as well as the number of Standard Model species emitted from an evaporating black hole~\cite{Page:1976df,MacGibbon:1991tj,Ray:2023auh}.
As the black hole temperature $\propto 1/M_{\rm BH}$, in the course of evaporation the number of particle species emitted, hence the Page factor, increases.
For $M_{\rm BH} > 10^{17}$~g, $P(M_{\rm BH}) = 1/1135 \pi$ with only photons and neutrinos emitted. 
For $M_{\rm BH} \ll 10^{10}$ g, $P(M_{\rm BH}) = 1/74 \pi$ with all Standard Model states emitted.
We use Ref.~\cite{MacGibbon:1991tj} to obtain the semi-analytical form of the Page factor, which agrees reasonably well with \texttt{BlackHawk}~\cite{BlackHawkArbey:2019mbc}.
In Eq.~\eqref{eq:BHevolution} we have conservatively omitted a growth rate term coming from ongoing dark matter accretion on the neutron star; it is generally negligible compared to the nucleon Bondi accretion rate, and the accretion of dark matter on the black hole is model-dependent~\cite{Garani:2018kkd}. 
We also neglect possible emission of non-standard states (such as $\chi$) from the nascent black hole, which can enhance evaporation rates and weaken our constraints. 

%%%%
\subsection{The effect of memory burden}
%%%%
The memory burden effect represents a deviation from standard Hawking radiation once a black hole loses a fraction $f_{\rm loss} \leq 1/2$ of its initial mass. 
Up to that point, the evaporation rate is described by Eq.~\eqref{eq:Hawkrate}.
Past that point, the black hole mass loss rate may be parameterized as~\cite{Dvali:2024hsb,thoss2024breakdownhawkingevaporationopens}
%%%%%%
\begin{equation}
 \dot{M}^{\rm memo}_{\rm evap} = \frac{1}{S(M_{\rm BH})^k} \frac{P(M_{\text{BH}})}{G^2 M_{\rm BH}^2}~,   
 \label{eq:MevapdotMB}
\end{equation}
%%%%%%
where $S(M_{\rm BH})= 4 \pi G M^2_{\rm BH}$ denotes the entropy of the black hole, and $k$ is restricted to positive integers in order to keep the lifetime of the black hole an analytic function of the entropy~\cite{Dvali:2024hsb}. 
Note that $k=0$ corresponds to the standard Hawking scenario.
In this work we will set $k=1$ to obtain results, but we find that these are practically unchanged for any $k\geq1$. 
This is because, in our parameter range of interest, the enormous entropy of the black hole (e.g., $S(10^{12}~{\rm g}) \sim 10^{34}$) practically switches off its evaporation in the memory-burdened phase. 
%Since the entropy is a very large number the $S(M_{\rm BH})^k$ term has the effect of suppressing evaporation. Due to this, any integer value of $k$ produces practically identical constraints as we will discuss below.
The memory burden effect may also precipitate a classical instability that triggers rapid decay of the black hole~\cite{Dvali:2020wft}, but we do not consider this possibility, and can be addressed in a future work.
%%%%
\section{Results}
\label{sec:results}
%%%%
In Fig.~\ref{fig:limits} we show our limits for spin-0 (left panel) and spin-1/2 (right panel) dark matter in the plane of spin-independent nucleon scattering cross section and dark matter mass.
In the top x-axis we mark the initial mass of the black hole corresponding to the $m_\chi$ in the bottom x-axis, as obtained from Eqs.~\eqref{eq:Mselfgrav} and \eqref{eq:MChandra}.
The shaded regions are excluded by the continued existence of PSR~J0437$-$4715. We also show limits from a recent search at the underground experiment LZ~\cite{LZCollaboration:2024lux}. 
Also shown are ``neutrino fog" cross sections~\cite{OHare:2021utq} below which dark matter direct detection is expected to be challenging due to an irreducible background from atmospheric neutrinos.
These limits are orders of magnitude weaker than our limits in the relevant mass range. 
This is simply because celestial objects, such as, neutron stars provide significantly larger target exposure to the incident dark matter flux. 
We note here, though, that direct detection limits are agnostic to the annihilation properties of dark matter, whereas we require negligible annihilations for black hole formation.

The mass of the nascent black hole scales as $M_{\rm BH} \propto m^{-3/2}_{\chi}$ for spin-0 and $M_{\rm BH} \propto m^{-2}_{\chi}$ for spin-1/2 dark matter (Eqs.~\eqref{eq:Mselfgrav} and \eqref{eq:MChandra}).
This implies, from Eqs.~\eqref{eq:BHevolution} and \eqref{eq:Hawkrate}, that as $m_\chi$ increases the black hole would require a longer time to consume the host star, and that Hawking evaporation becomes faster. 
These two effects ensure that above some $m_\chi$ the neutron star does not transmute into a black hole. 
For the survival of PSR~J0437$-$4715, requiring that the black hole's timescale to consume the neutron star be larger than the age of the neutron star, and that the Hawking evaporation timescale be smaller than the same, sets the following limits for spin-0 dark matter:
\begin{equation}
    m_\chi \geq \begin{cases}
         4.0 \times 10^9~{\rm GeV},~{\rm Bondi}~, \\
         7.5 \times 10^7~{\rm GeV},~{\rm Hawking}~.
    \end{cases}
    \label{eq:mxmaxspin0}
\end{equation}
 For spin-1/2 dark matter, these limits are 
\begin{equation}
    m_\chi \geq \begin{cases}
         3.2 \times 10^{11}~{\rm GeV},~{\rm Bondi}~, \\
         1.6 \times 10^{10}~{\rm GeV},~{\rm Hawking}~.
    \end{cases}
    \label{eq:mxmaxspin1/2}
\end{equation}
It is the limit from Hawking evaporation that sets the boundary on $m_\chi$ in constraints set by previous authors: the blue vertical lines in Fig.~\ref{fig:limits}.
The limit from Bondi accretion is shown in Fig.~\ref{fig:limits} as gray vertical lines.

We revise former limits by extending them into the red shaded region, where suppressed evaporation allows lighter black holes to consume the neutron star.
We have taken the fraction of mass lost in the Hawking stage of evaporation $f_{\rm loss}$ as 1/2, and taken the suppression exponent $k$ in Eq.~\eqref{eq:MevapdotMB} as 1.
These choices already push the limit on $m_\chi$ to within a factor of 2 of the Bondi accretion limit in Eqs.~\eqref{eq:mxmaxspin0} and \eqref{eq:mxmaxspin1/2}.
Refs.~\cite{Dvali:2015aja,Michel:2023ydf} have shown that quantum backreaction could set in at an earlier time and suppress black hole evaporation.
In this light, we find that for smaller values of $f_{\rm loss}$, despite a longer period of suppressed evaporation from memory burden, our limits do not improve by much because we rapidly run into the Bondi accretion limit in Eqs.~\eqref{eq:mxmaxspin0} and \eqref{eq:mxmaxspin1/2}, i.e. the gray vertical line in Fig.~\ref{fig:limits}.
As mentioned before, we also find that setting $k$ to integer values $>1$ has no effect on the limits in Fig.~\ref{fig:limits} due to the entropy-dependent suppression in the evaporation rate in Eq.~\eqref{eq:MevapdotMB}. 

%%%%%%
\section{Discussion}
\label{sec:concs}
%%%%%%
In our study we have invoked the memory burden effect to rederive constraints on interactions of particle dark matter in scenarios where it forms black holes in the interiors of neutron stars.
Multiple interesting avenues of study await investigation. 
Small dark matter-seeded black holes inside carbon-oxygen white dwarfs may trigger Type Ia-like thermonuclear explosions by depositing energy from evaporation~\cite{Fedderke:2019jur,Acevedo:2019gre,Janish:2019nkk,Garani:2023esk}.
Black holes may also be formed by gradual dark matter accumulation inside other stellar objects~\cite{Acevedo:2020gro,Ray:2023auh,Bhattacharya:2024pmp}, which may alter the evolution of main sequence stars~\cite{Bellinger:2023wou}, and may have implications for recurrent cosmology~\cite{Bramante:2024idl}. 
The breakdown of Hawking evaporation would redraw the constraints estimated in these studies, especially relevant for stellar objects with colder core temperatures. 
One could also consider neutron stars that turn into black holes via bosonic dark matter that forms a Bose-Einstein condensate and/or has self-interactions.
We have assumed that the shape of the radiation spectrum is unaffected in the memory burden phase in order to parameterize the evaporation rate as Eq.~\eqref{eq:MevapdotMB}, but it might be interesting to study modifications to observables for other spectral shapes.

It is compelling that advances in quantum gravity may hold the key to uncovering the nature of dark matter.
%%%%%%
\section*{Acknowledgments}
%%%%%%
We thank Sulagna Bhattacharya, Basudeb Dasgupta, Joachim Kopp, Chethan Krishnan, Ranjan Laha, and Tao Xu for helpful conversations. A.R. acknowledges support from the National Science Foundation (Grant No. PHY-2020275) and to the Heising-Simons Foundation (Grant No. 2017- 228).

%%%%%%%%%%%%%%%%%%%%%%%%%%%%%%%%%%%%%%%%%%%%%%%%%%
% \bibliographystyle{JHEP}
\bibliography{refs}
%%%%%%%%%%%%%%%%%%%%%%%%%%%%%%%%%%%%%%%%%%%%%%%%%%
\end{document}